\documentclass[10pt,twocolumn,english]{revtex4}
\usepackage[T1]{fontenc}
\usepackage[latin9]{inputenc}
\usepackage[a4paper]{geometry}
\usepackage{amsmath}
\usepackage{amssymb}
\usepackage{graphicx}

\makeatletter
\@ifundefined{textcolor}{}
{%
 \definecolor{BLACK}{gray}{0}
 \definecolor{WHITE}{gray}{1}
 \definecolor{RED}{rgb}{1,0,0}
 \definecolor{GREEN}{rgb}{0,1,0}
 \definecolor{BLUE}{rgb}{0,0,1}
 \definecolor{CYAN}{cmyk}{1,0,0,0}
 \definecolor{MAGENTA}{cmyk}{0,1,0,0}
 \definecolor{YELLOW}{cmyk}{0,0,1,0}
}

\makeatother

\usepackage{babel}
\begin{document}

\title{$O(N)$--Universality Classes and the Mermin-Wagner Theorem}

\author{Alessandro Codello$^{a}$%
\thanks{codello@sissa.it%
} and Giulio D'Odorico$^{a,b}$%
\thanks{dodorico@sissa.it%
}\\
}

\address{$^{a}$SISSA, Via Bonomea 265, 34136 Trieste, Italy\\
$^{b}$INFN, Sezione di Trieste, Italy}
\begin{abstract}
We study how universality classes of $O(N)$--symmetric models depend
continuously on the dimension $d$ and the number of field components
$N$. We observe, from a renormalization group perspective, how the
 implications of the Mermin-Wagner-Hohenberg theorem set in as we
gradually deform theory space towards $d=2$.
For fractal dimension in the range $2<d<3$ we find, for any $N\geq1$, a finite family of multi-critical
effective potentials of increasing order. Apart for the $N=1$ case, these disappear in $d=2$ consistently with the Mermin-Wagner-Hohenberg theorem.
Finally, we study $O(N=0)$--universality classes and find an infinite family of these in two dimensions. 
\end{abstract}
\maketitle

\paragraph*{Introduction.}

Our modern understanding of quantum or statistical
field theory is based on the ideas put forward by K. Wilson
and formalized within the framework of the renormalization group (RG) \cite{Wilson}.
This approach considers all possible theories describing
the quantum or statistical fluctuations of a given set of degrees
of freedom, the fields, subject only to the constraints imposed
by symmetry and dimensionality; this defines what we call theory space.
The process of quantization on
one side, or averaging on the other, is then seen as a trajectory
connecting the bare action or Hamiltonian to the full quantum
or statistical effective action. This trajectory can be of finite or infinite length (with respect
to the RG time); in the first case one is performing
an effective field theory calculation, while in the second case one needs
an ending point for the trajectory: this usually is a fixed-point. RG
fixed-points describe scale invariant theories,
where fluctuations on all length scales are equally important:
these theories, like lighthouses, shed light on the structure of theory space.
They attract or repel surrounding theories giving rise to universality, a 
phenomenon that underlies both non-perturbative renormalization
and the understanding of continuous phase transitions \cite{Wilson}.
Once all fixed-points are known we can reconstruct the general (topological)
properties of the RG flow and acquire a deep understanding of a given class of models.
A paradigmatic example of this is the $c$-theorem \cite{Zamolodchikov}, which describes the
RG flow between two dimensional theories.

Important information about two dimensional theories comes
from exact results for particular lattice models; still, our ability to predict
the universal features of two dimensional continuous phase transitions resides on our
understanding of the structure of theory space.
Three dimensional systems are much more difficult to
treat exactly; here too, many analytical insights come from the RG study,
otherwise one would have to resort to numerical methods.
Deep insights, such as the role played by conformal symmetry in constraining
statistical fluctuations, are also naturally embedded in the larger framework
of RG analysis \cite{Codello_2012}.

In this Letter we show how another fundamental and broad
result like the Mermin-Wagner-Hohenberg theorem \cite{MWH1,MWH2},
which states that there cannot be continuous phase transitions in
$d=2$ systems characterized by continuos symmetries, fits in the
RG picture. We will do this by studying scalar $O(N)$--models,
a class of theories that has many applications to low dimensional systems: they can describe long polymer chains ($N=0$), liquid-vapor ($N=1$), superfluid helium ($N=2$), ferromagnetic ($N=3$) and QCD chiral ($N=4$) phase transitions \cite{Berges_Tetradis_Wetterich_2002,PV}.
Despite their relevance, there is no complete description of how universality
classes of $O(N)$--models depend continuously on both $d$ and $N$. In this Letter
we give such a description by studying scaling solutions of the effective
average action \cite{Berges_Tetradis_Wetterich_2002}.
As a result we find many new $N\geq2$ universality classes
describing multi-critical models in fractal dimension $2\leq d \leq3$.
In the $N=0$ case we observe an infinite number of fixed-points in $d=2$,
analogue to the $N=1$ minimal--models \cite{minmod}.\\

\paragraph*{Flow equations.}

The effective average action (EAA) $\Gamma_{k}[\varphi]$ is a functional
that depends on the infrared scale $k$ and that interpolates smoothly
between the bare action for $k\rightarrow\infty$ and the standard effective
action for $k\rightarrow0$  \cite{Berges_Tetradis_Wetterich_2002}. The
EAA satisfies an exact RG equation \cite{Wetterich_1993} that describes
its dependence upon changes of scale; this equation can be used to
set up a framework where to concretely implement the RG ideas discussed
above. It is generally quite difficult to follow exactly the flow
of the EAA and to find the relative fixed-point
functionals: approximations are needed. One that retains important
information about the structure of theory space is the one where all
one-particle-irreducible (1PI) vertices of the EAA are evaluated
at zero momenta. This defines the running effective potential $U_{k}(\rho)$
which is a function of the $O(N)$--invariant $\rho=\frac{1}{2}\varphi^{2}$.
In this approximation theory space is represented by the functional space of effective
potentials. This space is still infinite dimensional and, at
least at the qualitative level, $O(N)$--universality classes of
the full theory can be found by determining the relative scaling solutions.

In terms of the running dimensionless effective potential $\tilde{U}_{k}(\tilde{\rho})=k^{-d}U_{k}(\rho)$, with $\tilde{\rho}=k^{-(d-2+\eta)}\rho$,
a scaling solution $\partial_{t}\tilde{U}_{*}(\tilde{\rho})=0$ satisfies
the following ordinary differential equation \cite{Wetterich_1993}:
\begin{eqnarray}
-(d-2+\eta)\tilde{\rho}\,\tilde{U}_{*}'+d\,\tilde{U}_{*} & = & c_{d}(N-1)\frac{1-\frac{\eta}{d+2}}{1+\tilde{U}'_{*}}\nonumber \\
 &  & +c_{d}\frac{1-\frac{\eta}{d+2}}{1+\tilde{U}'_{*}+2\tilde{\rho}\,\tilde{U}_{*}''}\,,\label{1}
\end{eqnarray}
where $c_{d}^{-1}=(4\pi)^{d/2}\Gamma(d/2+1)$. The
anomalous dimension $\eta$ fixes the scaling properties of the field
at a particular fixed-point; to lowest order its value is related to the running dimensionless effective potential
by \cite{Berges_Tetradis_Wetterich_2002}:
\begin{equation}
\eta=c_{d}\frac{4\tilde{\rho}_{0}\tilde{U}_{*}''(\tilde{\rho}_{0})^{2}}{\left[1+2\tilde{\rho}_{0}\tilde{U}_{*}''(\tilde{\rho}_{0})\right]^{2}}\,,\label{2}
\end{equation}
with $\tilde{\rho}_{0}$ the absolute minimum $\tilde{U}_{*}'(\tilde{\rho}_{0})=0$.

Every scaling solution, together with its domain of attraction, represents
a different universality class; thus by finding the solutions of the
system composed of (\ref{1}) and (\ref{2}) one can determine
$O(N)$--universality classes. Differently from other implementations
of the RG, all the analysis can be made leaving $d$ and $N$ as
free parameters, permitting us to study how theory space depends on
these.
\begin{figure}
\begin{centering}
\includegraphics[scale=0.62]{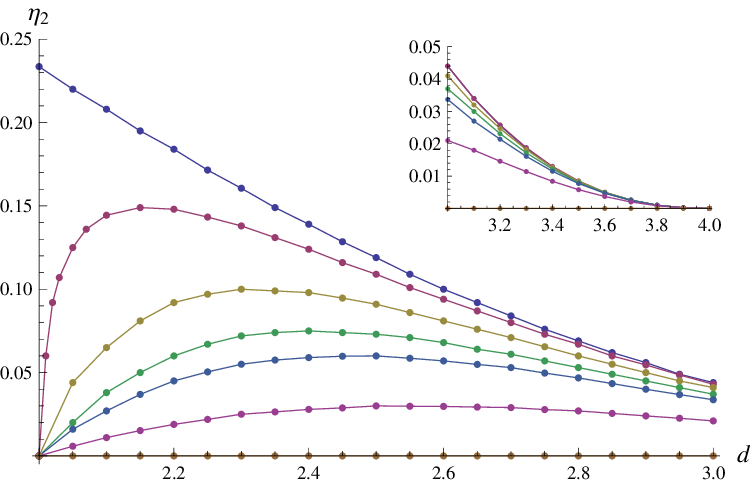}
\par
\end{centering}
\caption{$\eta_{2}$ as a function of $d$ for (from above)
$N=1,2,3,4,5,10,100$. In the inset we show the anomalous dimensions
in the range $3\leq d\leq4$ (note that the $N=1$ and $N=2$
curves are almost overlapping).}
\end{figure}
\\

\paragraph*{Mermin-Wagner-Hohenberg theorem.}

We solve the fixed-point equations (\ref{1}) and (\ref{2}) by the
iterative method proposed in \cite{Codello_2012}. For every $d$ and $N$ we find a discrete
set of scaling solutions to these equations. These correspond to multi-critical potentials of increasing order with $i$ minima (which we label by $i$), which are potentials describing multi-critical transitions, in which one needs to tune multiple parameters to reach the critical point. For each of these it is possible to obtain the anomalous dimension $\eta_{i}$ as a function of $d$ and $N$.
By studying the function $\eta_{i}(d,N)$  we can follow the evolution through theory space of the fixed-point representing the $i$--th multi-critical potential. 


\begin{figure}
\begin{centering}
\includegraphics[scale=0.62]{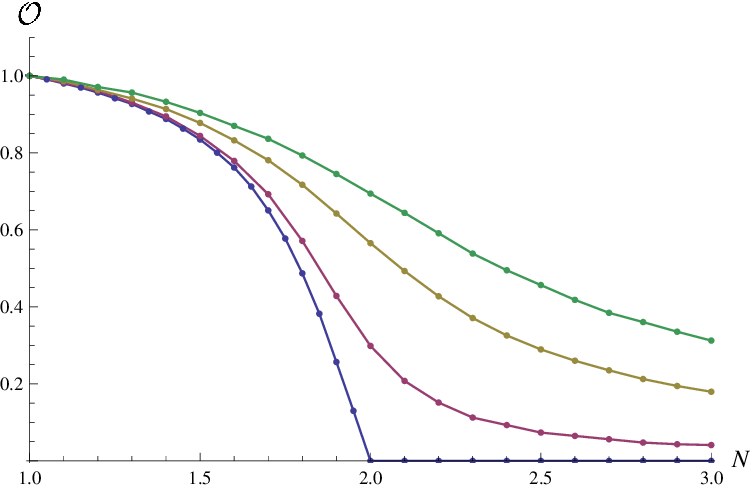}
\par
\end{centering}
\caption{$\mathcal{O}(d,N)=\eta_{2}(d,N)/\eta_{2}(d,1)$ as a function of $N$ for (from above) $d=2.1,2.05,2.01,2$.
$\mathcal{O}(d,N)$ can be interpreted as the order parameter of a
continuos phase transition in which $N$ plays the role of the control parameter.}
\end{figure}

For $d>4$ 
we find only the Gaussian fixed-point ($i=1$); at $d=4$, the upper critical
dimension for $O(N)$--models, the Wilson-Fisher fixed-points ($i=2$) start
to branch away from the Gaussian fixed-point.
In $d=3$ these fixed-points describe the known universality classes of the Ising, XY, Heisenberg
and other models; our estimates for the anomalous dimensions turn out to be
in good agreement with estimates available in the literature \cite{PV,Canet_Delamotte_Mouhanna_Vidal_2002}.   
Approaching $d=2$ one clearly observes that only
the $N=1$ anomalous dimension continues to grow
\cite{ftnote}: for all other values
of $N\geq2$ the anomalous dimension bends downward to become zero
exactly when $d=2$. This is a non-trivial fact, not evident from
the structure of equation (\ref{1}), telling us that only the
$O(N)$--model with discrete symmetry ($N=1$) can have a second-order
phase transition in two dimensions, while all the $O(N)$--models with
continuous symmetry ($N\ge2$) cannot.
This result, that here emerges solely from the RG analysis, is commonly
known as the Mermin-Wagner-Hohenberg (MWH) theorem \cite{MWH1}.
In this respect Figure 1
shows the way in which the MWH theorem manifests itself in the RG framework;
our analysis can be seen as a RG confirmation of this important
theorem and can be the starting point for a new rigorous proof of it.
Note also that, as expected from the exact solution  \cite{Morris:1997xj}, the anomalous dimension tends to zero for $N\rightarrow\infty$.

\begin{figure*}
\begin{centering}
\includegraphics[scale=0.44]{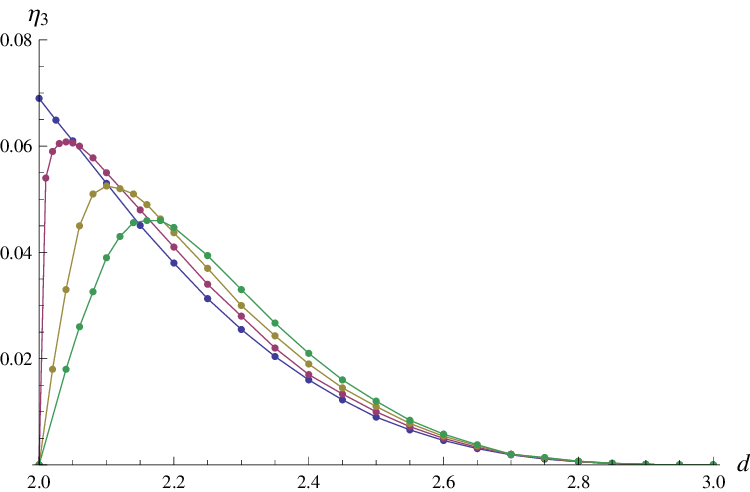}\includegraphics[scale=0.45]{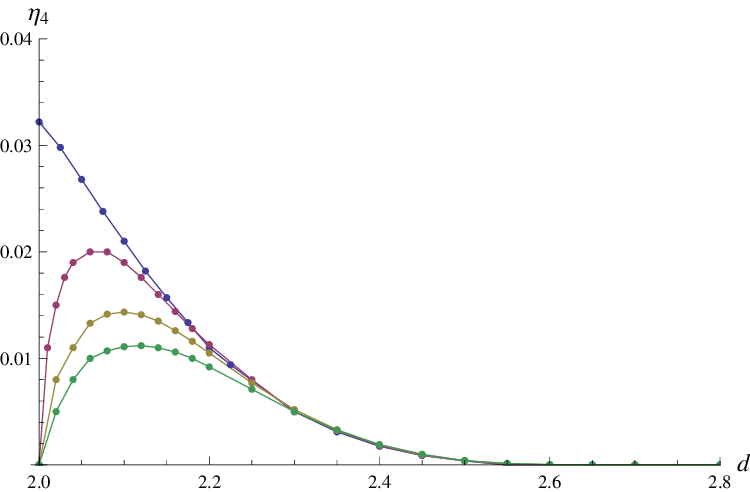}\includegraphics[scale=0.45]{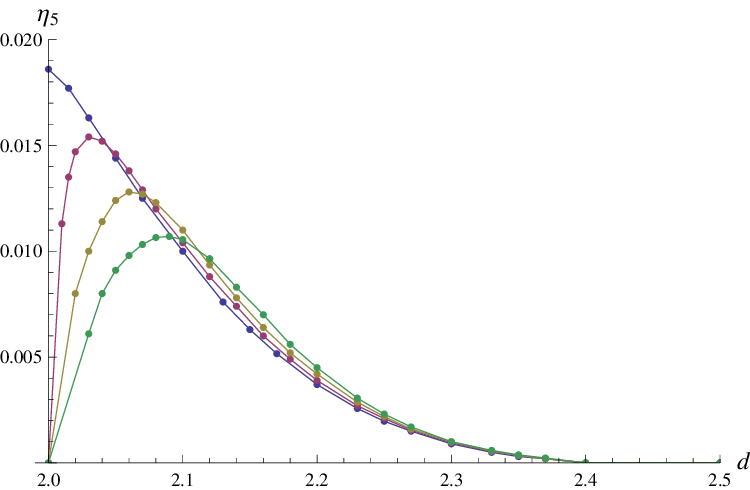}
\par
\end{centering}
\caption{$\eta_{i}$ as a function of $d$ for (from left)
the tri-critical ($i=3$), tetra-critical ($i=4$) and penta-critical ($i=5$) scaling solutions
for (from top at $d=2$) $N=1,2,3,4$.}
\end{figure*}

That the vanishing of the anomalous dimension implies that there are no continuous phase transitions for the $N\geq2$
models in $d=2$ can be confirmed by the analysis of the critical
exponent $\nu_{2}(d,N)$, which indeed blows up for $d\rightarrow2$ and $N\geq2$ \cite{CD}.
This allows us to distinguish the Spherical model, related to the $N\rightarrow\infty$ limit,
from the Gaussian model, both having $\eta=0$. Only
the $N=1$ model has a finite $\nu_{2}$ in two dimensions, in all other
cases $\nu_{2}$ diverges upon approaching $d=2$, as in the $N\rightarrow\infty$
limit where one knows exactly that $\nu_{2}(d,\infty)=\frac{1}{d-2}$.

The critical case $N=2$ is known to have a distinguished behavior \cite{Kosterlitz}. 
In this case one can observe all the distinctive properties of the
Kosterliz-Thouless phase transition by studying the properties 
of the RG flow \cite{Wetterich_XY}.

Our functions $\eta_2(d,N)$ can be compared with large--$N$ expansion
analogs \cite{Ma} which fail  to reproduce the small $N$ region, both qualitatively ($N=1$) and quantitatively ($N<10$).
To our knowledge, our method is the only able to give accurate theoretical estimates valid for every  $d$ and $N$.

To better discriminate between theories which can undergo a continuous phase transition in $d_{*}=2$
and those which cannot, we extend the analysis of scaling solutions
to non-integer $N$; in particular we want to see what
happens around the critical value $N_{*}=2$. The MWH theorem tells
us that at $d=d_{*}$ the quantity $\mathcal{O}(d,N)=\eta_{2}(d,N)/\eta_{2}(d,1)$
 can be seen as a sort of order parameter, meaning it
is zero for $N>N_{*}$ and non-zero for $N<N_{*}$; but it tells
us nothing about its  continuity in $N$.
Figure 2 shows that the RG analysis can say a lot more about this. First,
we see that $\mathcal{O}(d,N)$ evolves continuously with $N$ across
$N_{*}$; second, we see that $\mathcal{O}(d,N)$ can be written in
a scaling form around the transition point $(d_{*},N_{*})=(2,2)$;
in particular we can write the following scaling relation:
\begin{equation}
\mathcal{O}(d_{*},N)\sim\begin{cases}
\begin{array}{c}
\left(\frac{N_{*}-N}{N_{*}}\right)^{\Theta}\\
0
\end{array} & \begin{array}{c}
N\rightarrow N_{*}^{-}\\
N\rightarrow N_{*}^{+}
\end{array}\end{cases}\,,\label{3}
\end{equation}
where we introduced a new scaling exponent $\Theta$. A fit from the
data displayed in Figure 2 gives the estimate $\Theta\approx0.98$
which is quite close to one.
Relation (\ref{3}) tells us how theory space deforms as we vary the control
parameter $N$. An interesting question is if relation
(\ref{3}) is universal, in the sense that the value of $\Theta$
is independent of the details of the implementation of the RG procedure 
but rather describes an inner property of the set of theory spaces parametrized by $N$.
One can make a similar reasoning by
keeping $N$ fixed at $N_{*}$ and varying $d$ around $d_{*}$:
\begin{equation}
\mathcal{O}(d,N_{*})\sim\begin{cases}
\begin{array}{c}
0\\
\left(\frac{d-d_{*}}{d_{*}}\right)^{\frac{1}{\Delta}}
\end{array} & \begin{array}{c}
d\rightarrow d_{*}^{-}\\
d\rightarrow d_{*}^{+}
\end{array}\end{cases}\,;\label{4}
\end{equation}
where we introduced the new scaling exponent $\Delta$ and included the information, taken from \cite{Cassi_1992}, that $\eta_2$ remains
zero for $N\geq N_{*}$ and $d\leq d_{*}$. A fit from the data displayed
in Figure 1 gives the approximate value $\Delta\approx1.86$.
Finally, we found that equation (\ref{1}) has a discrete set of solutions only when the coefficient of the first term on the lhs is negative,
thus our analysis applies when $\eta>2-d$. This fact prevents us from performing
a complete analysis in the range $1\leq d<2$, where indeed studies of $O(N)$--models on
fractals have shown that the MWH theorem is still valid \cite{Cassi_1992}.
\\

\paragraph*{Multi-critical $O(N)$--models in fractal dimension.}

When new universality classes appear by branching from the Gaussian
fixed-point it is easy to determine the relative critical dimensions,
since the argument based on canonical dimensions is valid. In particular,
the $i$--th multi-critical scaling solution appears at the upper critical dimension $d_{c,i}=2+\frac{2}{i-1}$
\cite{Codello_2012}. At these dimensions we see non-trivial fixed-points branching from the Gaussian for every
$N\geq2$, corresponding to potentials with $i$ minima when expressed in terms of the variable $2\sqrt{\tilde{\rho}}$.

The critical dimensions $d_{c,i}$ accumulate at $d=2$ and thus one may naively expect to find, for any $N$, infinitely
many universality classes in two dimension. Our analysis shows instead,
see Figure 3 for the cases $i=3,4,5$ and $N=1,2,3,4$,
that this happens only in the $N=1$ case, where the multi-critical fixed-points approach,
in the limit $d\rightarrow2$, the fixed-points representing minimal-models \cite{Codello_2012}.
For any other $N\geq2$ we find that, consistently with the MWH theorem, the multi-critical
scaling solutions, present
in the range $2<d<3$, are instead absent in $d=2$. This fact is a strong check of
the general validity of the MWH theorem, which our analysis indicates is also applicable to multi-critical phase transitions.
On the other side, we predict the existence
of a whole family of $O(N)$--universality classes in fractal
dimensions between two and three. To our knowledge these universality
classes are new.\\


\paragraph*{The $N\rightarrow0$ limit.}

We now study the $N\rightarrow0$
limit, that describes the universality class of  self-avoiding random walks
(SAW) \cite{SAW}. Figure 4 (Top) shows $\eta_2$ as function
of $N$ in the interval between $-2\leq N\leq2.5$ for the cases $d=2$ and $d=3$.
The anomalous dimension is
continuous in the whole range; this is an indication that the $N\rightarrow0$
limit is well defined. Figure 4 (Top) also shows, interestingly,
that both the $d=2$ and $d=3$ curves tend to zero as $N\rightarrow-2$ where
indeed the model is know to have Gaussian critical exponents in both dimensions \cite{Fisher1973}.

We also find multi-critical scaling solutions for $N=0$. The interesting
thing here is that these solutions survive in infinite number
when $d\rightarrow2$.
A plot of the first four anomalous dimensions is shown in Figure 4 (Bottom);
these are numerically very similar to those of the $N=1$ models (see Figure
1 and 3).
This similarity is expected, as one may see by inspection of Figure 4 (Top).
Even if the anomalous dimension is
not a relevant physical parameter in the correspondence with SAW, we can use scaling relations to
relate it to the physical critical exponents $\nu$ and $\gamma$.
In $d=2$ one finds the exact values 
$\nu_{ex}=\frac{3}{4}$ and $\gamma_{ex}=\frac{43}{32}$ \cite{Vanderzande_1998}, 
and so $\eta_{ex}=2-\frac{\gamma_{ex}}{\nu_{ex}}=\frac{5}{24}\simeq0.208$;
we find $\eta_{2}(2,0)=0.232$. In $d=3$ one finds from Monte Carlo simulations the
values $\nu_{MC}=0.587$ and $\gamma_{MC}=1.157$ \cite{PV}, and
so $\eta_{MC}=2-\frac{\gamma_{MC}}{\nu_{MC}}\simeq0.029$; we find $\eta_{2}(3,0)=0.04$.
As we said before, we cannot extend our method to $d<2$ to compare
with exact SAW critical exponents found on fractals \cite{SAW on Fractals}.
In any case, our analysis suggests that there is a countable family
of $O(N=0)$--universality classes in two dimensions. To our knowledge
these are novel and may describe multi-critical phase transitions of some polymeric system.
\\
\begin{figure}
\begin{centering}
\includegraphics[scale= 0.60]{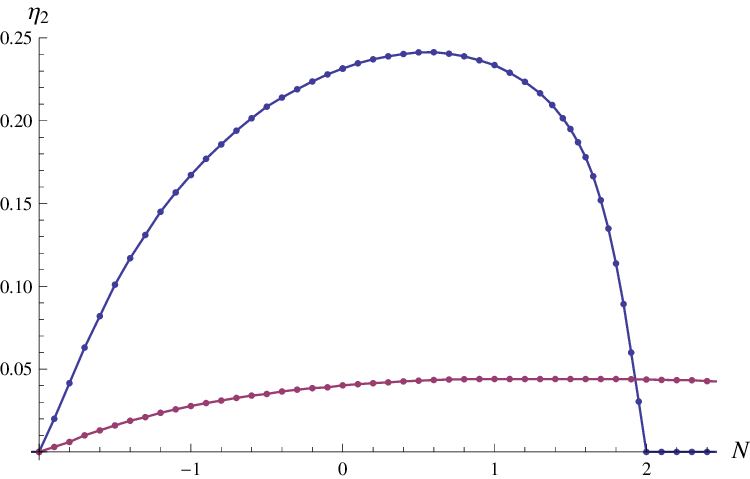}
\includegraphics[scale= 0.60]{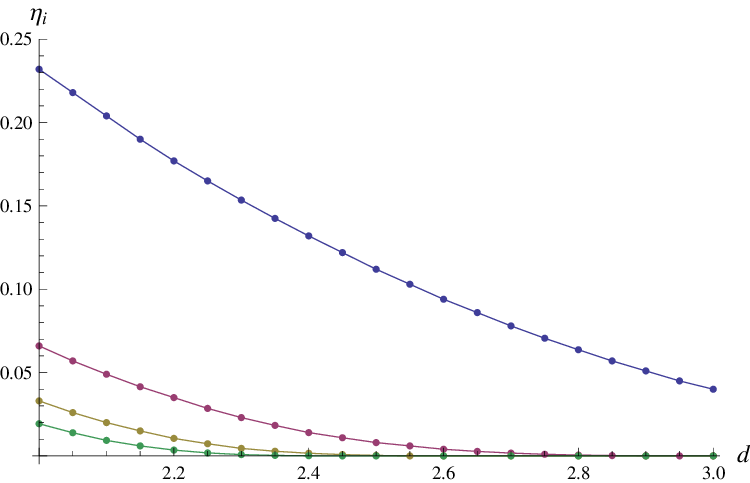}
\par
\end{centering}
\caption{(Top) $\eta_{2}$ as a function of $N$ is continuos both in $d=2$ (upper curve)
and in $d=3$ (lower curve). (Bottom) $\eta_{i}$ as a function of $d$ for the first four $N=0$ multi-critical scaling solutions, i.e. for (from above) $i=2,3,4,5$}
\end{figure}

\paragraph*{Discussion and outlook.}

In this Letter we studied how universality classes of scalar theories with linearly realized $O(N)$--symmetry
vary continuously with the dimension $d$ and with the number of field
components $N$. 
As we varied these parameters, we followed the evolution of RG fixed-points by
studying the scaling solutions of the RG equation (\ref{1}).
As in \cite{Codello_2012}, even if all our analysis was based on the study of a simple ODE, we
were able to observe a very rich behavior.

Above four dimensions, as expected, we found only the Gaussian universality
class; at $d=4$ we observed the Wilson-Fisher universality classes appear.
In fractal dimension between two and three we found non-trivial fixed-points for all $N$: these
are novel universality classes that can, in principle, be observed
in theoretical models on fractal lattices or in real physical systems.

Approaching two dimensions we observed the RG manifestation
of the MWH theorem: only the $N=1$ 
universality classes survived down to $d=2$, while all the $N\geq2$ ones disappeared.
By considering $(d,N)$ as real parameters near $(d_*,N_*)=(2,2)$ we found that the transition described by the MWH theorem, between
theories that can undergo a continuous phase transition and theories that cannot, is continuous,
and that the anomalous dimension, which can be seen as analogous to the order parameter, can be written in scaling form at the critical point $(2,2)$.
Our analysis revealed how different theory spaces parametrized by $N$ are related to each other; this information gives a deep RG understanding of the MWH theorem and could be used as the starting point for an extension of it. 

Finally, we studied the $N\rightarrow0$ limit;
we found that it is continuous around $N=0$ and we observed new $O(N=0)$--universality
classes in $d=2$. These are analogous to the  universality classes of $N=1$ minimal-models and may describe particular multi-critical transitions
of polymeric systems.



\begin{thebibliography}{10}

\bibitem{Wilson}K.G. Wilson, Rev. Mod. Phys. {\bf 47} (1975) 773;
K.G. Wilson, Rev. Mod. Phys. {\bf 55} (1983) 583.

\bibitem{Zamolodchikov}A.B. Zamolodchikov, JETP Lett. {\bf 43} (1986) 730, [Pisma Zh. Eksp. Teor. Fiz.  {\bf 43} (1986) 565].

\bibitem{Codello_2012} A. Codello,  arXiv:1204.3877 [hep-th].

\bibitem{MWH1}N. D. Mermin and H. Wagner . Phys. Rev. Lett. 17(22) (1966) 1133;
P. C. Hohenberg, Phys. Rev. 158(2) (1967);
Coleman, S. (1973) Comm. Math. Phys. {\bf 264} (30819) 259.

\bibitem{MWH2}A. Gelfert and W. Nolting, J. Phys. Con. Mat. {\bf 13} (2001) R505, cond-mat/0106090.

\bibitem{Berges_Tetradis_Wetterich_2002}J. Berges, N. Tetradis and C. Wetterich, Phys. Rept.  {\bf 363} (2002) 223, hep-ph/0005122.

\bibitem{PV}A. Pelissetto and E. Vicari, Phys. Rep. {\bf 368} (2002) 549.

\bibitem{minmod}A.B. Zamolodchikov, Sov. J. Nucl. Phys. {\bf 44} (1986), 529.

\bibitem{Wetterich_1993}C. Wetterich, Phys. Lett. B {\bf 301} (1993) 90.

\bibitem{Canet_Delamotte_Mouhanna_Vidal_2002}L. Canet, B. Delamotte,
D. Mouhanna and J. Vidal, Phys. Rev. D {\bf 67} (2003) 065004, hep-th/0211055;
L. Canet, B. Delamotte, D. Mouhanna and J. Vidal, Phys. Rev. B {\bf 68} (2003) 064421, hep-th/0302227.

\bibitem{ftnote}Our result $\eta_{2}(2,1)=0.234$  is in good agreement with the exact result $\eta_{ex}=0.25$; to provide an error on this estimate one needs to consider higher orders of the derivative expansion \cite{Canet_Delamotte_Mouhanna_Vidal_2002}.

\bibitem{Morris:1997xj}N. Tetradis and D.F.Litim, Nucl. Phys. B {\bf 464} (1996) 492, [hep-th/9512073];
T.R. Morris and M.D. Turner, Nucl. Phys. B {\bf 509} (1998) 637, [hep-th/9704202].

\bibitem{CD} A. Codello and G. D'Odorico, \emph{in preparation}.

\bibitem{Kosterlitz}J.M. Kosterlitz and D.J. Thouless, J. Phys. C {\bf 6} (1973) 1181.

\bibitem{Wetterich_XY}M. Grater and C. Wetterich, Phys. Rev. Lett.{\bf 75} (1995) 378, [hep-ph/9409459];
G. Von Gersdorff and C. Wetterich, Phys. Rev. B {\bf 64} (2001) 054513, [hep-th/0008114].

\bibitem{Ma}S. Ma, The 1/$n$ Expansion, in C. Domb and M.S. Green, Eds., "Phase Transitions and Critical Phenomena" Vol. 6 (1976) Accademic Press. 

\bibitem{Cassi_1992}D. Cassi, Phys. Rev. Lett. {\bf 68} (1992) 3631.

\bibitem{SAW}De Gennes, Phys. Lett. A {\bf 38} (1972) 339.

\bibitem{Fisher1973}M. Fisher, Phys. Rev. Lett. {\bf 30} (1973) 679;
R. Balian,and G. Toulouse, Phys. Rev. Lett. {\bf 30}  (1973) 544.

\bibitem{Vanderzande_1998}B. Nienhuis, J. Stat. Phys. {\bf 34} (1984) 731.

\bibitem{SAW on Fractals}S. Elezovic, M. Knezevic and S. Milosevic, J. Phys. A {\bf 20} (1987) 1215.


\end{thebibliography}
\end{document}